\pdfoutput=1

\documentclass[aps,10pt,twocolumn,amsmath,amssymb,preprintnumbers,floatfix,prb,superscriptaddress,longbibliography]{revtex4-2}

\usepackage[utf8]{inputenc}
\usepackage{newtxtext}
\usepackage[upint]{newtxmath}
\usepackage{microtype}
\usepackage{textcomp}
\usepackage{eucal}
\usepackage{bm}
\usepackage{siunitx}
\usepackage{comment}
\usepackage{lipsum}

\usepackage{enumerate}
\usepackage{amsfonts}
\usepackage{amsmath}
\usepackage{amssymb}
\usepackage{color}
\usepackage{soul}

\usepackage{graphicx}

\usepackage[colorlinks,allcolors=blue]{hyperref}
\usepackage[capitalize]{cleveref}

\definecolor{DarkRed}{rgb}{0.65,0,0}%
\definecolor{Green}{rgb}{0,0.3,0.3}
\definecolor{Purple}{rgb}{0.3,0,0.65}
\definecolor{Red}{rgb}{1,0,0}
\definecolor{Blue}{rgb}{0,0,0.85}
\definecolor{Magenta}{rgb}{1,0,1}

\newcommand{\ve}[1]{\boldsymbol{#1}}

\newcommand{\vech}{\ve{h}} 

\newcommand{\veck}{\ve{k}}

\newcommand{\rkky}{\text{RKKY}}
\newcommand{\fermi}{\text{F}}
\newcommand{\Tr}{\text{Tr}}

\newcommand{\be}{\begin{equation}}
\newcommand{\ee}{\end{equation}}

\newcommand{\prlsection}[1]{\textit{#1}.\kern0.05em---\kern0.05em\ignorespaces}

\begin{document}
\title{RKKY interaction in Rashba altermagnets}
\author{Morten Amundsen}
\affiliation{Center for Quantum Spintronics, Department of Physics, Norwegian \\ University of Science and Technology, NO-7491 Trondheim, Norway}
\author{Arne Brataas}
\affiliation{Center for Quantum Spintronics, Department of Physics, Norwegian \\ University of Science and Technology, NO-7491 Trondheim, Norway}
\author{Jacob Linder}
\affiliation{Center for Quantum Spintronics, Department of Physics, Norwegian \\ University of Science and Technology, NO-7491 Trondheim, Norway}

\begin{abstract}
The interaction between two impurity spins provides vital information about the host system and has been suggested to form a building block in quantum computation and spintronic devices. We here determine this Ruderman–Kittel–Kasuya–Yosida (RKKY) interaction in recently discovered altermagnetic materials, including the presence of a Zeeman field, in a two- and three-dimensional altermagnet. In two dimensions, we also study the effect of Rashba spin-orbit coupling. Our results reveal that the momentum-resolved spin-polarization of the itinerant carriers in the altermagnet changes the RKKY interaction qualitatively from the isotropic spin-splitting in a ferromagnet. The Ising-contribution directed parallel with the altermagnetism is found to exhibit a beating pattern reflecting the shape of the Fermi surface of the system and may thus be further influenced by an out-of-plane Zeeman field. However, the exchange interaction for in-plane impurity spins deviates only slightly from that of a normal metal. On shorter length scales, we find that the Ising, the Dzyaloshinskii-Moriya terms, and other non-collinear interaction terms in the RKKY interaction acquire a rapidly oscillating behavior as a function of the relative angle between the impurity spins, which is not present in the ferromagnetic case. We determine how this new length scale depends on the system parameters analytically. Our results show that the RKKY interaction in altermagnets is qualitatively different from that of ferromagnets despite both breaking time-reversal symmetry.
\end{abstract}
\maketitle

\section{Introduction} 
When impurities are introduced to a material, they disturb their surroundings like ripples in a pond, owing to the wavelike nature of quantum mechanical particles. Through these fluctuations, pairs of spin-polarized impurities may communicate in a rather peculiar way, alternating between favoring either a parallel or antiparallel alignment of their spin, depending on the distance. This is known as the Ruderman–Kittel–Kasuya–Yosida (RKKY) interaction\cite{kasuya1956,ruderman1954,yosida1957}. The nature of this interaction is closely related to the spin susceptibility of the surrounding material and, therefore, provides information about its properties. In a non-magnetic, spin-degenerate material (which we will refer to as a normal metal), the spin impurities exhibit a behavior describable by the Heisenberg model: collinear alignment of spins without any preferential direction. Such oscillations have been observed both in layered structures~\cite{majkrzak1986,grunberg1986,parkin1990,parkin1991}, as well as between magnetic adatoms on material surfaces~\cite{zhou2010,khajetoorians2012,pruser2014}. 
\begin{figure}[t!]
\includegraphics{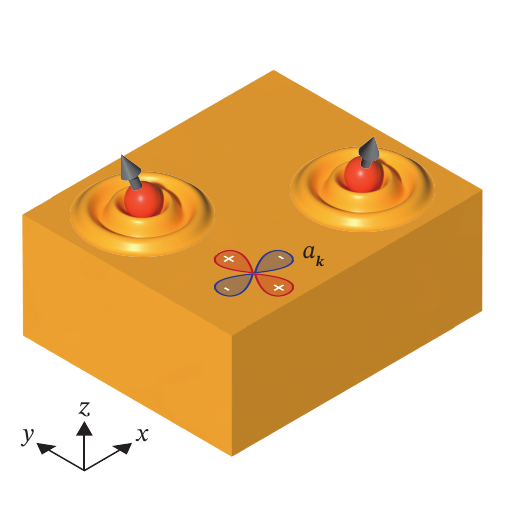}
\caption{The system considered. Spin impurities are introduced to an altermagnet, characterized by a spin splitting $a_{\bm{k}} = \hbar^2\beta(k_x^2-k_y^2)/2m$. We also consider the effects of a Zeeman field and a spin-orbit coupling. The impurities influence each other via the RKKY interaction.}
\label{fig:geometry}
\end{figure}
In spin-polarized materials, an $XXZ$-type exchange interaction is typically found instead---still a collinear alignment between impurity spins, but with different oscillatory behavior for spins aligned parallel with the spin polarization~\cite{parhizgar2013,valizadeh2015,valizadeh2016}. In materials that break inversion symmetry and therefore feature significant spin-orbit coupling, non-collinear alignment may also appear~\cite{smith1976,fert1980,zhu2011,bouaziz2017,wang2017}, typically in the form of a Dzyaloshinskii-Moriya (DM) interaction~\cite{dzyaloshinsky_jpcs_58,moriya_pr_60}, a finding which is also well-established experimentally~\cite{grigoriev2008,grigoriev2010,dupe2015,khajetoorians2016}. The preference for a non-collinear alignment can be understood from the fact that the electron spin precesses when moving through the momentum-dependent exchange field provided by the spin-orbit coupling.

Since the RKKY interaction is so closely connected with the properties of the material through which it is mediated, it may be used as a probe to gain insight into the inner workings of complex materials. Here, we predict the form of the RKKY interaction in altermagnets~\cite{hayami2019,smejkal2020,yuan2020,mazin2021,smejkal2022}, which is a class of magnetic materials that are neither ferromagnetic nor antiferromagnetic but displays properties conventionally associated with both. It has no net magnetization, but its dispersion features a momentum-dependent spin splitting, having a $d$-wave form $\sim k_x^2-k_y^2$. A range of theoretical and experimental works have already fleshed out much of the interesting physics that takes place in altermagnets, such as its spin transport properties~\cite{sun2023,zhang,lyu2024}, its topological properties~\cite{fernandes2024} as well as its interaction with superconductors~\cite{li2023,cheng2024,zhu2023,papaj2023,beenakker2023,sun2023,ouassou2023, brekke_prb_23}. Here, we investigate how spin-polarized impurities behave as they interact with each other through a two- or three-dimensional altermagnet. Our study corroborates the findings of Ref.~\cite{Lee2024}, which considers the RKKY interaction in a two-dimensional pure altermagnet, and goes beyond by also considering the simultaneous presence of Rashba spin-orbit coupling and a Zeeman field. We find that the RKKY oscillations feature a beating pattern that is highly directional in nature. Moreover, we find that the RKKY interaction promoting collinear spin coupling and the DM interaction arising when Rashba spin-orbit coupling is present both oscillate rapidly at a fixed separation distance of the impurities when varying only the direction of their separation vector.

\section{Theory} 
\begin{figure*}[t!]
\includegraphics{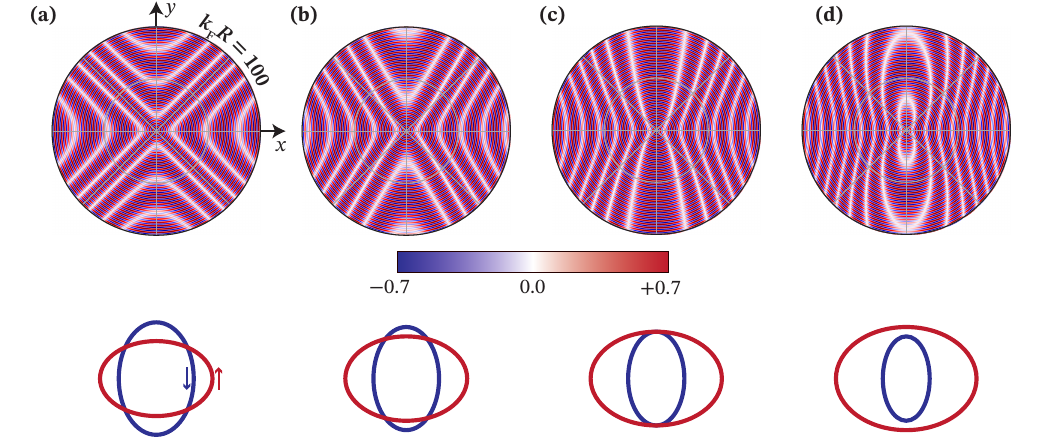}
\caption{The beating pattern induced in the Ising component $J_{zz}(\bm{R})$ of the RKKY interaction by the altermagnetism in two dimensions, and its evolution in the presence of an increasing $z$ component of the Zeeman field for (a) $h_z/\mu = 0$, (b) $h_z/\mu = 0.5\beta$, (c) $h_z/\mu = \beta$, and (d) $h_z/\mu = 1.5\beta$. The results in three dimensions are similar. The insets beneath each figure show the corresponding dispersion relation at the Fermi level. We have here chosen $\beta = 0.1$. For clarity, we have multiplied the exchange interaction, given in \cref{eq:Jzz}, with $(k_\fermi R)^2$ to compensate for the decay.}
\label{fig:beating}
\end{figure*}

The system is in \cref{fig:geometry}. Spin impurities are placed in the $xy$ plane of an altermagnet which is either two- or three-dimensional. The Hamiltonian is given by $H = \sum_{\veck} H(\bm{k})$, with
\begin{align}
H(\bm{k}) = \xi_k\sigma_0 + a_{\veck}\sigma_z - \left[\alpha(\hat{z}\times\bm{k})+\bm{h}\right]\cdot\bm{\sigma}
\end{align}
where $\xi_k = \hbar^2k^2/2m - \mu$, and $a_{\veck} = \hbar^2\beta(k_x^2-k_y^2)/2m$, with $\beta$ the (dimensionless) strength of the altermagnetism, satisfying $0 < \beta < 1$ and $k=|\veck|$ the magnitude of the crystal momentum, which is either two- or three-dimensional. The parameter $\alpha$ indicates the strength of the Rashba spin-orbit coupling, a term which we include only in the analysis of a two-dimensional system, where it can generated for instance via a gate voltage. The parameter $\bm{h}$ is a Zeeman field, which may be induced by an applied magnetic field, or via proximity to a ferromagnetic insulator. Furthermore, $\sigma_0$ is the $2\times2$ identity matrix, and $\sigma_j$ for $j\in\left\{x,y,z\right\}$ are the Pauli matrices. 

We place several spin impurities onto the $xy$ plane of this system, at positions $\bm{r}_j$, thereby introducing a perturbation of the form
\begin{align}
H' = \frac{J}{N}\sum_j\sum_{\veck\veck'} e^{i(\bm{k} - \bm{k}')\cdot\bm{r}_j}\bm{S}_j\cdot \bm{\sigma},
\end{align}
where $J$ is the exchange interaction, and $N$ the number of particles in the system. To linear order in $J$, this perturbation gives a contribution proportional to $\bm{h}$ from the coupling between the spin impurity and the spin density of the substrate. While this coupling dominates the RKKY interaction, we disregard it based on the following. This contribution is spatially constant and can be eliminated by first determining the exchange energy of a single impurity, e.g. by spin-polarized scanning-tunneling microscopy (STM) measurements ~\cite{ghanbari2021}. 

To second order in $J$, the zero temperature RKKY interaction between a pair of spins, $\bm{S}_1$ and $\bm{S}_2$ is given as
\begin{align}
H_\rkky = -\frac{J^2}{\pi}\text{Im}\int_{-\infty}^{0}d\varepsilon\; \Tr\left[\left(\bm{S}_1\cdot\bm{\sigma}\right) G(\bm{R})\left(\bm{S}_2\cdot\bm{\sigma}\right)G(-\bm{R})\right],
\label{eq:rkky}
\end{align}
where $\bm{R} = \bm{r}_1 - \bm{r}_2$ is the distance between the impurities and $G(\bm{R},\varepsilon)$ is the retarded Green function of the altermagnet substrate. In momentum space, it is given as $G(\bm{k},\varepsilon) = \left[ \varepsilon - H + i\delta\right]^{-1}$, for infinitesimal $\delta$, which becomes
\begin{align}
G(\bm{k}) = \frac{1}{D}\left(\vphantom{\frac{1}{2}}  (\varepsilon-\xi_k)\sigma_0 + a_{\veck}\sigma_z -\left[\alpha(\hat{z}\times\bm{k})+\bm{h}\right]\cdot\bm{\sigma} \right),
\end{align}
with 
\begin{align*}
D&=(\varepsilon-\xi_k+i\delta)^2 - (a_{\veck}-h_z)^2\\ 
&- (\alpha k_y-h_x)^2- (\alpha k_x + h_y)^2. 
\end{align*}
If we further assume that $\alpha$, $h_x$, and $h_y$ are small and include their effects only to the first order, we get
\begin{align}
G(\bm{k}) &\simeq \begin{pmatrix} g_+(\bm{k}) & 0 \\ 0 & g_-(\bm{k})\end{pmatrix}\nonumber\\
&-\left[\alpha(\hat{z}\times\bm{k})+\bm{h}_\perp\right]\cdot\bm{\sigma} g_+(\bm{k})g_-(\bm{k}),
\label{eq:Gk}
\end{align}
where $\bm{h}_\perp = h_x\hat{x} + h_y\hat{y}$. The functions $g_\pm(\bm{k})$ are the Green functions of the altermagnet in the absence of both impurities and spin-orbit coupling and are given as
\begin{align}
g_{s}(\bm{k}) &= \frac{1}{\varepsilon - \xi_{ks} + i\delta},
\label{eq:gs}
\end{align}
with $\xi_{ks} = \xi_k + s(a_{\veck}-h_z)$. In position space, \cref{eq:Gk} takes the form
\begin{align}
G(\bm{R}) &\simeq G_0(\bm{R})\sigma_0 + \bm{G}(\bm{R})\cdot\bm{\sigma},
\label{eq:GR}
\end{align}
where
\begin{align*}
G_0(\bm{R}) &= [g_+(\bm{R})+g_-(\bm{R})]/2, \\
G_x(\bm{R}) &= -(h_x +i\alpha\partial_y)Q(\bm{R}), \\
G_y(\bm{R}) &= -(h_y -i\alpha\partial_x)Q(\bm{R}), \\
G_z(\bm{R}) &= [g_+(\bm{R})-g_-(\bm{R})]/2.
\end{align*}
Furthermore, the position-space version of $g_s(\bm{k})$ is found to be
\begin{align*}
g^{(n)}_s(\bm{R}) = \int \frac{d\bm{k}}{(2\pi)^n}\; g_s(\bm{k}) e^{i\bm{k}\cdot\bm{R}}=-\frac{m\xi_n(\rho_s)}{\hbar^2\sqrt{1-\beta^2}}e^{ic_s\rho_s},
\end{align*}
where $n\in \{2,3\}$ is the dimension of the substrate, $c^2_s = 2m(\varepsilon +\mu+sh+ i \delta)/\hbar^2$, and $\rho_s^2 = \rho_{s,x}^2 + \rho_{s,y}^2$ in two dimensions, and  $\rho_s^2 = \rho_{s,x}^2 + \rho_{s,y}^2+ \rho_{s,z}^2$ in three dimensions, with $\rho_{s,x} = R_x\sqrt{1+s\beta}$, $\rho_{s,y} = R_y\sqrt{1-s\beta}$, and $\rho_{s,z} = R_z$. The function $\xi_n$ is given as
\begin{align*}
\xi_n(\rho_s) = \begin{cases} \frac{i}{\sqrt{2i\pi c_s\rho_s}}, & n = 2 \\ \frac{1}{2\pi\rho_s}, & n = 3 \end{cases}
\end{align*}
We note that the expression for $g_s^{(3)}$ is exact, whereas the one for $g_s^{(2)}$ is valid in the limit $c_sR\gg 1$.
Finally, $Q(\bm{R})$, which is the inverse Fourier transform of $Q(\bm{k}) = g_+(\bm{k})g_-(\bm{k})$, is given as
\begin{align*}
Q(\bm{R}) = \int \frac{d\bm{k}}{(2\pi)^n}\; \frac{e^{i\bm{k}\cdot\bm{R}}}{\left(\varepsilon - \xi_{k+}+i\delta\right)\left(\varepsilon - \xi_{k-} + i\delta\right)}.
\end{align*}
By Feynman parametrization, this can be rewritten as
\begin{align}
Q(\bm{R}) = \frac{1}{2}\int_{-1}^{1}du\int \frac{d\bm{k}}{(2\pi)^n}\; \frac{e^{i\bm{k}\cdot\bm{R}}}{\left(\varepsilon - \tilde{\xi}_{k+}+i\delta\right)^2},
\end{align}
with $2m\tilde{\xi}_{k+}/\hbar^2 = (1+u\beta)k_x^2 + (1-u\beta)k_y^2 + k_z^2+\mu+uh$. Performing the momentum integral results in
\begin{align}
Q(\bm{R}) = -\frac{m^2}{2\hbar^4}\int_{-1}^{1}du\frac{\rho_+^u\xi_n(\rho_+^u)e^{ic^u_+\rho_+^u}}{ic^u_+\sqrt{1-u^2\beta^2}},
\end{align}
with $\rho_+^u$ ($c_+^u$) defined in the same way as $\rho_+$ ($c_+$), but with $\beta\to u\beta$ ($h\to uh$). An asymptotic expression for $Q(\mathbf{R})$, valid in the limit $c_\pm R\gg1$ may be found to leading order by integrating by parts,
\begin{align}
Q(\mathbf{R})&\simeq-\frac{m^2}{\hbar^4\sqrt{1-\beta^2}} \frac{1}{\beta(R_x^2-R_y^2)}\times\nonumber\\
&\sum_{s=\pm1}s\left(\frac{\rho_s}{ic_s}\right)^2\xi_n(\rho_s)e^{ic_s\rho_s}.
\end{align}
An important point is that only the last term in \cref{eq:GR}, due to the spin-orbit coupling, is odd under inversion $\bm{R}\to-\bm{R}$. By inserting \cref{eq:GR} into \cref{eq:rkky} we find
\begin{align}
H_\rkky = -\frac{2J^2}{\pi}\text{Im}\int_{-\infty}^0 d\varepsilon\;&\left[\left(G_0^2-\bm{G}\cdot\tilde{\bm{G}}\right)(\mathbf{S}_1\cdot\mathbf{S}_2)\right.\nonumber\\
&+ iG_0(\tilde{\mathbf{G}}-\mathbf{G})\cdot(\mathbf{S}_1\times\mathbf{S}_2) \nonumber\\
&+ (\mathbf{S}_1\cdot\mathbf{G})(\mathbf{S}_2\cdot\tilde{\mathbf{G}})\nonumber\\
&+\left.(\mathbf{S}_1\cdot\tilde{\mathbf{G}})(\mathbf{S}_2\cdot\mathbf{G})\vphantom{\left(G_0^2\right)}\right],
\end{align}
with $\tilde{X}(\bm{R}) = X(-\bm{R})$. This produces an RKKY interaction of the form 
\begin{align}
H_\rkky &= J_{zz}(\bm{R})S_1^zS_2^z + J_{xx}(\bm{R})\left(S_1^xS_2^x+S_1^yS_2^y\right) \nonumber\\
&+J_{zn}(\bm{R})\left(S_{1}^zS_2^n + S_1^nS_2^z\right) + \bm{D}(\bm{R})\cdot\left(\bm{S}_1\times\bm{S}_2\right).
\label{eq:heff}
\end{align}
\cref{eq:heff} has a form typical of any material with a spin-splitting and inversion symmetry breaking, such as topological insulators and semiconductors with spin-orbit coupling. The effect of the altermagnet appears in the interaction parameters and their dependence on the separation distance $\bm{R}$ between the impurities. The RKKY interaction in a pure altermagnet, when only $J_{zz}$ and $J_{xx}$ are present, has a $C_4$ symmetry owing to the $d$-wave form of the altermagnet. Furthermore, the spin splitting along the $z$ axis breaks spin rotational symmetry, with the consequence that $J_{zz}$ and $J_{xx}$ become distinct. When the Zeeman field $\bm{h}$ is introduced, the $C_4$ symmetry is broken, reducing to a $C_2$ symmetry. In addition, when $\bm{h}$ is not parallel with the altermagnetism, it will introduce a non-collinear coupling $J_{zn}$ between the impurity spins. With spin-orbit coupling, we lose the $C_2$ symmetry, because of the inversion symmetry breaking, thereby generating a DM interaction $\bm{D}$. We will in the following discuss results in the long-distance regime $k_\fermi R \gg 1$, with a selection of analytically exact results, valid at all length scales, given in \cref{appendix1}. The Ising contribution $J_{zz}$ becomes
\begin{align}
J_{zz}(\bm{R}) &= \frac{J_n}{4}\sum_{s=\pm1}\frac{\sin \left(2k_s\rho_s - \frac{n\pi}{2}\right)}{\left(k_\fermi\rho_s\right)^n}
\label{eq:Jzz}
\end{align}
with $k_s = k_\fermi \sqrt{1+sh_z/\mu}$, where we defined $k_\fermi^2 = 2m\mu/\hbar^2$, which is the Fermi wave vector of a normal metal, and $J_n = k_\fermi^{2n}J^2/(2\pi)^n\mu(1-\beta^2)$.

The in-plane contribution $J_{xx}$, stemming from the interaction between impurity spins aligned in the plane orthogonal to the altermagnet spin splitting direction, is given as

\begin{align}
J_{xx}(\bm{R}) =4J_n
\left(\frac{1-\left(\frac{h_z}{\mu}\right)^2}{k_+\rho_+k_-\rho_-}\right)^{\frac{n-1}{2}}\frac{\sin \left(2\overline{k\rho} - \frac{n\pi}{2}\right)}{k_-\rho_++k_+\rho_-},
\label{eq:Jxy}
\end{align}
where $\overline{k\rho} = (k_+\rho_++k_-\rho_-)/2$.

In the following, when discussing in-plane magnetization $\bm{h}_\perp$ or spin-orbit coupling, we will assume that the out-of-plane component of the magnetization is zero, $h_z = 0$. To linear order in $\bm{h}_\perp$, we obtain
\begin{align}
J_{zn}(\bm{R}) &= \frac{\bm{h}_\perp\cdot\hat{n} J_n}{4\mu}J_h(\bm{R}) 
\label{eq:Jzn}
\end{align}
with 
\begin{align*}
J_h(\bm{R}) &= \frac{1}{k_\fermi^2(\rho_+^2-\rho_-^2)}\left[\sum_{s=\pm1}(k_\fermi\rho_s)^{2-n}\sin\left(2k_\fermi\rho_s + \frac{n\pi}{2}\right)\right.\\
&\left.- \frac{2R^2}{\bar{\rho}(\rho_+\rho_-)^{\frac{n-1}{2}}}\sin \left(2k_\fermi\bar{\rho}+\frac{n\pi}{2}\right)\right],
\end{align*}

Finally, the Rashba spin-orbit coupling gives rise to a DM interaction of the form
\begin{align}
D_j(\mathbf{R}) &= \frac{D_0R_j}{(\rho_+^2-\rho_-^2)}\text{Im}\sum_s s\Gamma_s\left[ e^{2ik_F\rho_s} + \frac{\rho_s^2}{\bar{\rho}\sqrt{\rho_+\rho_-}}e^{2ik_F\bar{\rho}}\right],
\label{eq:DMI}
\end{align}
with $\eta_x = -1$, $\eta_y = 1$, $D_0 =2\alpha\eta_jJ_2/\mu$ and
\begin{align*}
\Gamma_s(\mathbf{r}) = \frac{4\eta_j\beta}{k_F^2(\rho_+^2-\rho_-^2)} + \frac{3(1-s\eta_j\beta)}{2k_F^2\rho_s^2} + \frac{i(1-s\eta_j\beta)}{k_F\rho_s}.
\end{align*}

\section{Results and Discussion}
We will here discuss key findings from the derivations in the previous section. Our findings are qualitatively similar in both two- and three-dimensional systems, and we restrict our discussion of results to two dimensions in the following, as this is the most experimentally feasible system. Indeed, the orientation of the spin of each individual spin impurity may then be probed via spin-polarized STM measurements~\cite{delgado2010,loth2010,zhou2010}.  We also note that, in three dimensions, the RKKY interaction reduces to that of a normal metal along the $z$ axis.

\subsection{Altermagnet with and without out-of-plane spin splitting}
We begin by exploring the large-scale patterns that emerge when the altermagnetism is combined with an out-of-plane Zeeman field $h_z$. It is known that the RKKY interaction exhibits a beating pattern when the dispersion of the underlying substrate features a spin splitting. This is because the two spin species of the mediating quasiparticles have slightly different Fermi momenta, giving a characteristic sinusoidal modulation with wavelength inversely proportional to the spin splitting. This is observed also in our system, as shown in \cref{fig:beating}, which presents results for the two-dimensional system. In three dimensions, the behavior is qualitatively similar. In (a), the beating pattern solely due to the altermagnetism is shown, with $\beta = 0.1$, as determined from \cref{eq:Jzz}. We multiplied this expression by a factor $(k_\fermi R)^2$ to compensate for the decay and display the beating pattern more clearly. The spin splitting forms nodal lines in the much more rapidly oscillating RKKY interaction, the distance between which is determined by size of the spin-splitting along a specific direction. The nodal lines occur most frequently along the $x$ and $y$ axes, where the spin-splitting is greatest, whereas no beating is observed along the diagonals $|x| = |y|$, reducing to normal-metal behavior. In (b), a Zeeman field of $h_z = 0.5\beta$ is applied, and so the beating wavelength along the $x$ axis is reduced because the spin-splitting here is increased, while the opposite is the case for the $y$ axis. The splitting of the two spin bands still cancels at a certain angle in momentum space, now different from $45^\circ$, which is observable as the absence of beating for interactions along the same angle in position space. This angle reaches $90^\circ$ when $h = \beta$, as shown in $(c)$, and we obtain a behavior in which the system behaves as a normal metal along the $y$ axis and as a ferromagnet along the $x$ axis. Finally, in (d), where $h = 1.5\beta$, the spin-splitting cancels nowhere, and beating is found regardless of the angular placement of the spin impurities.  
\begin{figure}[t!]
\includegraphics{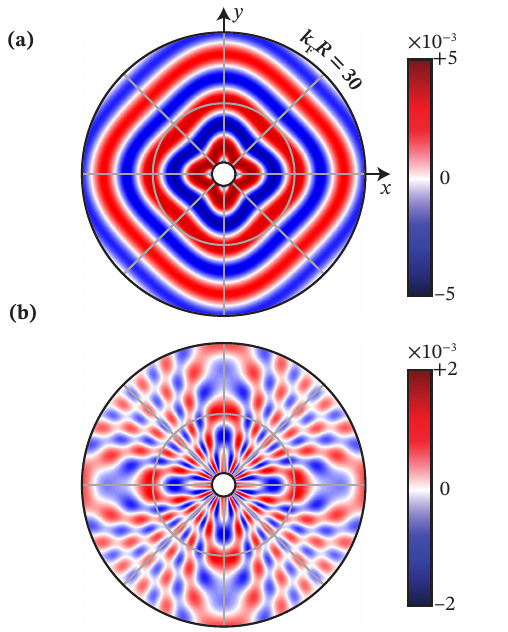}
\caption{The position dependence of the exchange interaction, with the strength of the altermagnetism set to $\beta  =0.5$, in the two-dimensional system. In three dimensions, the results are qualitatively similar. In \textbf{(a)}, $J_{xx}(\mathbf{R})/J_2$ is shown, whereas \textbf{(b)} is showing $J_{zz}(\bm{R})/J_2$, as defined in \cref{eq:Jxy} and \cref{eq:Jzz}, respectively. The radial coordinate indicates the distance between a pair of impurities, $k_\fermi R\in \left[20,30\right]$. As the altermagnetic strength $\beta$ increases, the RKKY interaction acquires a strongly oscillatory behavior as one varies the direction of the separation vector of the spin impurities even at a fixed distance (following a circular contour in the above plots). This is different from both normal metals and ferromagnets.}
\label{fig:AM}
\end{figure}
\begin{figure}[t!]
\includegraphics{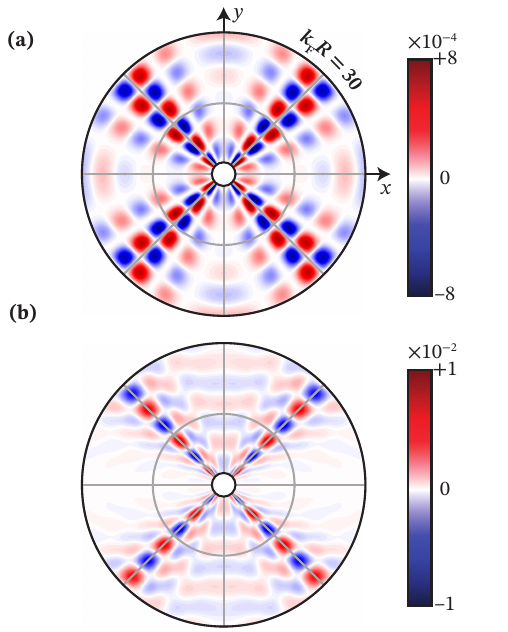}
\caption{The spatial dependence of (a) $J_{zn}(\mathbf{R})/J_2$, the exchange interaction produced by an exchange field $\bm{h}_\perp$ directed in the plane orthogonal to the altermagnetism, defined in \cref{eq:Jzn}, and (b) the $x$ component $D_x(\bm{R})/ J_2$ of the Dzyaloshinskii-Moriya interaction generated by the spin-orbit coupling, as defined in \cref{eq:DMI}. The $y$ component is found by a $90^\circ$ rotation of $D_x$. The results are here shown for a two-dimensional system. Here, we have set $|\bm{h}_\perp|/\mu = \alpha k_\fermi/\mu = 0.1$. The choice of parameters is otherwise as in \cref{fig:AM}. }
\label{fig:SOC}
\end{figure}

Next, we consider the interaction between spins oriented in the plane orthogonal to the spin-splitting induced 
by the altermagnetism and $h_z$, namely $J_{xx}$, as given in \cref{eq:Jxy}. In that case, the interaction between the spin impurities, whose spin are now orthogonal to the spin-splitting direction of the substrate, is mediated by spin-flip processes of the quasiparticles, hybridizing the two spin bands. We thus no longer obtain a superposition of the interaction generated by the two spin species but something much more reminiscent of an average, whereby the effect of the spin splitting is significantly reduced. Indeed, for $\beta\ll 1$ and $h_z/\mu\ll1$, the RKKY interaction reduces to that of a normal metal, with no beating present. Even larger values for $\beta$ have only a modest effect on $J_{xx}$, as shown in \cref{fig:AM}(a) for $\beta = 0.5$, $h_z = 0$, and $k_\fermi R\in[20,30]$, where it is seen that the interaction deviates only slightly from that of a normal metal, which has radial symmetry. For comparison, $J_{zz}$ is shown in the same parameter regime in \cref{fig:AM}(b), where the RKKY oscillations are seen to be influenced to a much greater degree, acquiring the $d$-wave-like form of the altermagnet dispersion. This has the interesting effect of creating angular oscillations in the spin interactions of a pair of spin impurities, meaning that the two spins oscillate between parallel and antiparallel configurations as one moves one impurity around the other, keeping their distance fixed. Note that this is a much shorter length scale than in \cref{fig:beating}, so any beating is not visible. These results are in agreement with the findings of Ref.~\onlinecite{Lee2024}.

\subsection{Altermagnet with in-plane spin splitting}
We now move on to an in-plane Zeeman field $\bm{h}_\perp$, and set $h_z = 0$ in the following. Considering first impurities placed along one of the diagonals of the altermagnet, $|x| = |y|$, it is straightforward to deduce the resulting behavior. This is a node for the altermagnetism, and hence, the system reduces to a ferromagnet, for which exact analytical expressions for the interaction terms in \cref{eq:heff} may be found by setting the spin quantization axis along $\bm{h}_\perp$, making this the distinguished axis in the spin-split RKKY interaction. However, no such global choice of quantization axis is possible for general placement of the spin impurities. Spin is not a good quantum number, and non-collinear interaction between impurity spins is possible. To linear order in $\bm{h}_\perp/\mu$, the resulting exchange interaction is given in \cref{eq:Jzn}, and shown in \cref{fig:SOC}(a) for $|\bm{h}_\perp|/\mu = 0.1$. The first thing to notice is that $J_{zn}$ vanishes when $|x| = |y|$, which perfectly corroborates the above explanation: non-collinear spin interaction is impossible along the nodal lines of the altermagnetism. Furthermore, a closer inspection of \cref{eq:Jzn} reveals that it results from a competition between interference of the two spin species and hybridization. Indeed, \cref{eq:Jzn} contains one contribution which depends on $k_\fermi\rho_s$ separately, which is typical for the former behavior (see, e.g., \cref{eq:Jzz}), and one term which depends on $k_\fermi\bar\rho$, which is indicative of the latter type of behavior (as in \cref{eq:Jxy}). On the diagonals, they cancel exactly.

\subsection{Altermagnet with Rashba spin-orbit coupling}
Lastly, we consider the effect of spin-orbit coupling to linear order in $\alpha$. We consider a two-dimensional material, with the SOC induced, e.g., by a gate voltage. This gives rise to a DMI, an expression for which is given in \cref{eq:DMI}. We here set $\vech=0$. Its $x$ component, $D_x$, is shown in \cref{fig:SOC}(b) for the same parameter set as in \cref{fig:AM}, and with $\alpha k_\fermi / \mu = 0.1$. The $y$ component $D_y$ is found by a $90^\circ$ rotation of the figure. We can understand its effect using a similar analysis for $\bm{h}_\perp$. The DMI also produces a non-collinear interaction between impurity spins. Still, whereas $J_{zn}$ is symmetric upon interchange of spin indices, $\bm{D}$ is antisymmetric, giving rise to the characteristic cross-product relationship $\bm{S}_1\times\bm{S}_2$. This is because spin-orbit coupling breaks inversion symmetry, or, said differently, the momentum-depending effective exchange field induced by the SOC flips sign when the direction is reversed. Hence, the same principles that dictate the influence of $\bm{h}_\perp$ apply also to the DMI. It, too, is a competition between interference and hybridization, but with the effective exchange field direction rotating with the direction of motion. In this case, there is no cancellation on the diagonals since the spin interaction of the substrate is never diagonal, even in the absence of altermagnetism. We also point out that $D_x$ vanishes along the $x$ axis for symmetry reasons dictated by the DMI vector selection rules \cite{dzyaloshinsky_jpcs_58,moriya_pr_60}.

\section{Conclusion}
Here, we have discussed the RKKY interaction in altermagnets in the presence of other spin-dependent fields. The altermagnet exhibits an angularly dependent beating pattern in the Ising contribution of the RKKY interaction, which a collinearly applied exchange field may influence. For impurity spins in the plane orthogonal to this spin splitting, however, the interaction shows only a slight deviation from a normal metal. We have also studied the effect of an exchange field $\bm{h}_\perp$ orthogonal to the altermagnetism and find that the spatial distribution of the exchange interaction consists of a competition between interference- and hybridization-dominated effects, with the former producing a much more anisotropic response than the latter. Finally, we study the impact of spin-orbit coupling and the resulting Dzyaloshinskii-Moriya interaction and explain its behavior by comparing it with the interaction generated by $\bm{h}_\perp$. 

 \begin{acknowledgments}
This work was supported by the Research
Council of Norway through Grant No. 323766 and its Centres
of Excellence funding scheme Grant No. 262633 “QuSpin.” Support from
Sigma2 - the National Infrastructure for High Performance
Computing and Data Storage in Norway, project NN9577K, is acknowledged.
 \end{acknowledgments}
 
\appendix
\section{Exact expressions}
\label{appendix1}
For a selection of cases, it is possible to derive exact analytical expressions for the RKKY interactions given in \cref{eq:heff}, valid at all length scales. We provide them below.

\subsection*{$J_{zz}$ in three dimensions for $\alpha = 0$, $\mathbf{h}_\perp = 0$}
\begin{align*}
J_{zz}(\mathbf{R}) = -\frac{J_3}{2}\sum_{s=\pm1}\frac{\sin 2k_s\rho_s -2k_s\rho_s\cos 2k_s\rho_s }{(k_\fermi\rho_s)^4}
\end{align*}

\subsection*{$J_{zz}$ in two dimensions for $\alpha = 0$, $\mathbf{h}_\perp = 0$}
\begin{align*}
&J_{zz}(\mathbf{R}) = \frac{J_2}{\pi}\times\\
&\sum_{s=\pm1}\left(1+s\frac{h_z}{\mu}\right)\left[J_0(k_s\rho_s)Y_0(k_s\rho_s) + J_1(k_s\rho_s)Y_1(k_s\rho_s)\right],
\end{align*}
where $J_n(x)$ and $Y_n(x)$ are the $n$'th order Bessel functions of the first and second kind, respectively.
\subsection*{$J_{xx}$ in three dimensions for $\alpha = 0$, $\mathbf{h} = 0$}
\begin{align*}
J_{xx}(\mathbf{R}) = -4J_3\frac{\sin 2k_\fermi\bar\rho -2k_\fermi\bar\rho\cos 2k_\fermi\bar\rho }{k_\fermi^4\rho_+^2\rho_-^2\bar\rho^2}.
\end{align*}
\subsection*{$J_{xx}$ in two dimensions for $\alpha = 0$, $\mathbf{h} = 0$}
\begin{align*}
&J_{xx}(\mathbf{R}) = \frac{J_2}{\pi}\frac{1}{\rho_+^2-\rho_-^2}\times\\
&\sum_{s=\pm1}s\rho_s\left[J_0(k_\fermi\rho_{-s})Y_1(k_\fermi\rho_s) + Y_0(k_\fermi\rho_{-s})J_0(k_s\rho_s)\right].
\end{align*}

\bibliography{masterref,AMRKKY}

\end{document}